# Lévy Measure Decompositions for the Beta and Gamma Processes

Yingjian Wang                                                                                       YW65@DUKE.EDU
Lawrence Carin                                                                                     LCARIN@DUKE.EDU
Electrical and Computer Engineering Department, Duke University, Durham NC 27708

## Abstract

We develop new representations for the Lévy measures of the beta and gamma processes. These representations are manifested in terms of an infinite sum of well-behaved (proper) beta and gamma distributions. Further, we demonstrate how these infinite sums may be truncated in practice, and explicitly characterize truncation errors. We also perform an analysis of the characteristics of posterior distributions, based on the proposed decompositions. The decompositions provide new insights into the beta and gamma processes (and their generalizations), and we demonstrate how the proposed representation unifies some properties of the two. This paper is meant to provide a rigorous foundation for and new perspectives on Lévy processes, as these are of increasing importance in machine learning.

## 1. Introduction

A prominent distinction of nonparametric methods relative to parametric approaches is the utilization of stochastic *processes* rather than probability *distributions*. For example, a Gaussian process (Rasmussen & Williams, 2006) may be employed to nonparametrically represent general smooth functions on a continuous space of covariates (*e.g.*, time). Recently the idea of nonparametric methods has extended to feature learning and data clustering, with interest respectively in the beta-Bernoulli process (Thibaux & Jordan, 2007) and the Dirichlet process (Ferguson, 1973). In such processes the nonparametric aspect concerns the number of features/clusters, which are allowed to be unbounded ("infinite"), permitting the model to adapt the number of these entities as the given and future data indicate. The increasing importance of these models in machine learning warrants a detailed theoretical analysis of their properties, as well as simple constructions for their implementation. In this paper we focus on Lévy processes (Sato, 1999), which are of increasing interest in machine learning.

A family of Lévy processes, the pure-jump nondecreasing Lévy processes, also fit into the category of the completely random measure proposed by Kingman (Kingman, 1967). The beta process (Hjort, 1990) is an example of such a process, which is applied in nonparametric feature learning. The gamma process falls in this family as well, with its normalization the well-known Dirichlet process. Hierarchical forms of such models have become increasingly popular in machine learning (Teh et al., 2006; Teh, 2006; Thibaux & Jordan, 2007), as have nested models (Blei et al., 2010), and models that introduce covariate dependence (MacEachern, 1999; Williamson et al., 2010; Lin et al., 2010).

As a consequence of the important role these models are playing in machine learning, there is a need for the study of the properties of pure-jump nondecreasing Lévy processes. As examples of such work, (Thibaux & Jordan, 2007) and (Paisley et al., 2010) present explicit constructions for generating the beta process, (Teh et al., 2007) derives a construction for the Indian buffet process parallel to the stick-breaking construction of the Dirichlet process (Sethuraman, 1994), and (Thibaux, 2008) obtains a construction for the gamma process under the gamma-Poisson context. Apart from these specialized construction methods, in (Kingman, 1967) a general construction method for completely random measures is proposed, by first decomposing it into a sum of a countable number of $\sigma$-finite measures, and then superposing the Poisson processes according to these sub-measures. By regarding the completely random measure as a Lévy process, this method corresponds to decomposing the Lévy measure, which provides clarity of theoretical properties and simplicity in practical implementation. However this Lévy measure decomposition method has not yet come into wide use in machine learning and statistics, probably due to the nonexistence of a universal construction of the measure decomposition.

In this paper we develop explicit and simple decompositions by following the conjugacy principle for two widely used Lévy processes, the beta and gamma processes. The conjugacy means that the decompositions are manifested by leveraging the forms of conjugate likelihoods to the Lévy measures. The decompositions bring new perspectives on the beta and gamma processes, with associated properties analyzed here in detail. The decompositions are constituted in terms of an infinite set of sub-processes of form convenient for computation. Since the number of sub-processes is infinite, a truncation analysis is also presented, of interest for practical use. We show some posterior properties of such decompositions, with the beta process as an example. We also extend the decomposition to the *symmetric* gamma process (positive and negative jumps), suggesting that the Lévy measure decomposition is applicable for other pure-jump Lévy processes represented by their Lévy measures. Summarizing the main contributions of the paper:

- We constitute Lévy measure decompositions for the beta, stable-beta, gamma, generalized gamma and symmetric gamma processes via the principle of conjugacy, providing new perspectives on these processes.

- The decomposition of the beta process unifies the constructions in (Thibaux & Jordan, 2007), (Teh & Görür, 2009), and (with a different decomposing method) (Paisley et al., 2010), and a new generative construction for the gamma process and its variations is derived.

- Truncation analyses and posterior properties for such decompositions are presented for practical use.

## 2. Background

Lévy processes (Sato, 1999) and completely random measures (Kingman, 1967) are two closely related concepts. Specifically, some Lévy processes can be regarded as completely random measures. In this section brief reviews and connections are presented for these two important concepts.

### 2.1. Lévy process

A Lévy process $X(\omega)$ is a stochastic process with independent increments on a measure space $(\Omega, \mathcal{F})$. $\Omega$ is usually taken to be one-dimensional, such as the real line, to represent a stochastic process with variation over time. By the Lévy-Itô decomposition (Sato, 1999), a Lévy process can be decomposed into a continuous Brownian motion with drift, and a discrete part of a pure-jump process. When a Lévy process $X(\omega)$ only has the discrete part and its jumps are positive, then for $\forall \mathcal{A} \in \mathcal{F}$ the characteristic function of the random variable $X(\mathcal{A})$ is given by:

$$\mathbb{E}\{e^{juX(\mathcal{A})}\} = \exp\{\int_{\mathbb{R}^+ \times \mathcal{A}} (e^{jup} - 1)\nu(dp, d\omega)\} \quad (1)$$

with $\nu$ satisfying the integrability condition (Sato, 1999). The expression in (1) defines a category of pure-jump nondecreasing Lévy processes, including most of the Lévy processes currently used in nonparametric Bayesian methods, such as the beta, gamma, Bernoulli, and negative binomial processes. With (1), such a Lévy process can be regarded as a Poisson point process on the product space $\mathbb{R}^+ \times \Omega$ with the mean measure $\nu$, called the Lévy measure. On the other hand, if the increments of $X(\omega)$ on any measurable set $\mathcal{A} \in \mathcal{F}$ are regarded as a random measure assigned on the set, then $X(\omega)$ is also a completely random measure. Due to this equivalence, in the following discussion we will not discriminate the pure-jump nondecreasing Lévy process $X$ with its corresponding completely random measure $\Phi$.

### 2.2. Completely random measure

A random measure $\Phi$ on a measure space $(\Omega, \mathcal{F})$ is termed "completely random" if for any disjoint sets $\mathcal{A}_1, \mathcal{A}_2 \in \mathcal{F}$ the random variables $\Phi(\mathcal{A}_1)$ and $\Phi(\mathcal{A}_2)$ are independent. A completely random measure $\Phi$ can be split into three independent components:

$$\Phi = \Phi_f + \Phi_d + \Phi_o \quad (2)$$

where $\Phi_f = \sum_{\omega \in \mathcal{I}} \phi(\omega)\delta_\omega$ is the fixed component, with the atoms in $\mathcal{I}$ fixed and the *jump* $\phi(\omega)$ random; $\mathcal{I}$ is a countable set in $\mathcal{F}$. The deterministic component $\Phi_d$ is a deterministic measure on $(\Omega, \mathcal{F})$. $\Phi_f$ and $\Phi_d$ are relatively less interesting compared to the third component $\Phi_o$, which is called the ordinary component of $\Phi$. According to (Kingman, 1967), $\Phi_o$ is discrete with both random atoms and jumps.

In (Kingman, 1967), it is noted that $\Phi_o$ can be further split into a countable number of independent parts:

$$\Phi_o = \sum_k \Phi_k, \quad \Phi_k = \sum_{(\phi(\omega), \omega) \in \Pi_k} \phi(\omega)\delta_\omega \quad (3)$$

Denote $\nu$ as the Lévy measure of (the Lévy process corresponding to) $\Phi_o$, $\nu_k$ as the Lévy measure of $\Phi_k$,

$\Pi$ a Poisson process with $\nu$ its mean measure, and $\Pi_k$ a Poisson process with $\nu_k$ its mean measure; (3) further yields:

$$\nu = \sum_k \nu_k, \quad \Pi = \bigcup_k \Pi_k \qquad (4)$$

which provides a constructive method for $\Phi_o$: first construct the Poisson process $\Pi_k$ underlying $\Phi_k$, and then with the superposition theorem (Kingman, 1993) the union of $\Pi_k$ will be a realization of $\Phi_o$. In the following sections we show how this general construction method of (4) can be applied on pure-jump nondecreasing Lévy processes of increasing interest in machine learning, with an emphasis on the beta and gamma processes, and their generalizations.

## 3. Beta process

A beta process (Hjort, 1990) is a Lévy process with beta-distributed increments; $B \sim \mathrm{BP}(c(\omega), \mu)$ is a beta process if

$$B(d\omega) \sim \mathrm{Beta}(c(\omega)\mu(d\omega), c(\omega)(1-\mu(d\omega))) \qquad (5)$$

where $\mu$ is the base measure on measure space $(\Omega, \mathcal{F})$ and a positive function $c(\omega)$ the concentration function. Expression (5) indicates that the increments of the beta process are independent, which makes it a special case of the Lévy process family. The Lévy measure of the beta process is

$$\nu(d\pi, d\omega) = c(\omega)\pi^{-1}(1-\pi)^{c(\omega)-1}d\pi\mu(d\omega) \qquad (6)$$

where $\mathrm{Beta}(0, c(\omega)) = c(\omega)\pi^{-1}(1-\pi)^{c(\omega)-1}$ is an *improper* beta distribution since its integral over $(0,1)$ is infinite. As a result, its *underlying Poisson process*, *i.e.*, the Poisson process with $\nu$ as its mean measure on the product space $\Omega \times (0,1)$, denoted $\Pi$, has an infinite number of points drawn from $\nu$, yielding

$$B = \sum_{i=1}^{\infty} \pi_i \delta_{\omega_i} \qquad (7)$$

where $\pi_i$ is the jump (increment) which happens at the atom $\omega_i$. Real variable $\gamma = \mu(\Omega)$ is termed the mass parameter of $B$, and we assume $\gamma < \infty$.

### 3.1. Beta process Lévy measure decomposition

The infinite integral of the improper beta distribution inspires a decomposition of the improper distribution with an infinite number of *proper* distributions. The singularity in the improper beta distribution is manifested from $\pi^{-1}$. Since $\pi \in (0,1)$, the geometric series expansion yields

$$\pi^{-1} = \sum_{k=0}^{\infty}(1-\pi)^k, \quad \pi \in (0,1) \qquad (8)$$

and substituting (8) in (6), with manipulation detailed in the Supplementary Material, we have the Lévy measure decomposition theorem of the beta process:

**Theorem 1** *For a beta process $B \sim \mathrm{BP}(c(\omega), \mu)$ with base measure $\mu$ and concentration $c(\omega)$, denote $\Pi$ as its underlying Poisson process and $\nu$ the Lévy measure, then $B$ and $\Pi$ can be expressed as*

$$\Pi = \bigcup_{k=0}^{\infty} \Pi_k, \quad B = \sum_{k=0}^{\infty} B_k \qquad (9)$$

*where $B_k$ is a Lévy process with $\Pi_k$ its underlying Poisson process. The Lévy measure $\nu_k$ of $B_k$ is a decomposition of $\nu$:*

$$\nu = \sum_{k=0}^{\infty} \nu_k$$
$$\nu_k(d\pi, d\omega) = \mathrm{Beta}(1, c(\omega)+k)d\pi\mu_k(d\omega) \qquad (10)$$
$$\mu_k(d\omega) = \frac{c(\omega)}{c(\omega)+k}\mu(d\omega)$$

*where $\mathrm{Beta}(1, c(\omega)+k)$ is the PDF of beta distribution with parameters $1$ and $c(\omega)+k$.*

Theorem 1 is the beta process instantiation of the completely random measure decomposing in (4), which indicates that the underlying Poisson process $\Pi$ of the beta process $B$ is the superposition of an infinite number of independent Poisson processes $\{\Pi_k\}_{k=0}^{\infty}$, with $\nu_k$ the mean measure of $\Pi_k$ and $\mu_k$ the mean measure of the restriction of $\Pi_k$ on $\Omega$. As a result, the beta process $B$ can be expressed as a sum of an infinite number of independent Lévy processes $\{B_k\}_{k=0}^{\infty}$ with $\{\Pi_k\}_{k=0}^{\infty}$ the underlying Poisson process. The independence of $\{\Pi_k\}_{k=0}^{\infty}$ and $\{B_k\}_{k=0}^{\infty}$ w.r.t. index $k$ is justified by the fact that both $\mu$ and $c(\omega)$ are fixed parameters.

### 3.2. The Lévy process $B_k$

It is interesting to study the properties of $B_k$, such as the expectation and variance. Denoting $\mathcal{B}_k(d\omega) = \frac{1}{c(\omega)+k+1}\mu_k(d\omega)$ as the base measure of $B_k$, for $\forall \mathcal{A} \in \mathcal{F}$:

$$\mathbb{E}(B_k(\mathcal{A})) = \int_{\mathcal{A}} \mathcal{B}_k(d\omega) = \mathcal{B}_k(\mathcal{A})$$
$$\mathrm{Var}(B_k(\mathcal{A})) = \int_{\mathcal{A}} \frac{2}{c(\omega)+k+2}\mathcal{B}_k(d\omega) \qquad (11)$$

It is noteworthy that the Lévy process $B_k$ is no longer a beta process, since (5) is not satisfied. By Theorem 1, the jumps of $B_k$ follow a *proper* beta distribution parameterized by the concentration function $c(\omega)$ and the index $k$, and $\mu_k$ determines the locations where the jumps happen. Since $\{B_k\}_{k=0}^{\infty}$ are independent w.r.t.

the index $k$, with Theorem 1:

$$\sum_{k=0}^{\infty} \mathbb{E}(B_k(\mathcal{A})) = \mathbb{E}(B(\mathcal{A})) \\ \sum_{k=0}^{\infty} \text{Var}(B_k(\mathcal{A})) = \text{Var}(B(\mathcal{A})) \quad (12)$$

The detailed procedure to derive (11) and (12) is given in the Supplementary Material.

### 3.3. Simulating the beta process

#### 3.3.1. Poisson superposition simulation

Theorem 1 reveals that the underlying Poisson process of a beta process is a superposition of an infinite number of Poisson processes, each of which has a *finite* set of atoms. This perspective also provides a simulation procedure for the beta process: first, the Poisson process $\Pi_k$ is sampled for all $k = 0, 1, 2, \cdots$, (here we term the index $k$ as the "round" of the simulation); then take the union of the samples of each $\Pi_k$ as a realization of the Poisson process $\Pi$. With the marking theorem (Kingman, 1993) implicitly applied, the simulation procedure of the beta process is as follows:

**Simulation procedure**: For round $k$:

1: Sample the number of points for $\Pi_k$: $n_k \sim \text{Poisson}(\int_\Omega \mu_k(d\omega))$;
2: Sample $n_k$ points from $\mu_k$: $\omega_{ki} \overset{\text{i.i.d.}}{\sim} \frac{\mu_k}{\int_\Omega \mu_k(d\omega)}$, for $i = 1, 2, \cdots, n_k$;
3: Sample $B_k(\omega_{ki}) \overset{\text{i.i.d.}}{\sim} \text{Beta}(1, c(\omega_{ki}) + k)$, for $i = 1, 2, \cdots, n_k$;

Then the union $\bigcup_{k=0}^{\infty} \{(\omega_{ki}, B_k(\omega_{ki})\}_{i=1}^{n_k}$ is a realization of $\Pi$ (and equivalently of $B$).

We refer to the above simulation procedure as the *Poisson superposition simulation*, for the central role of the Poisson superposition. The especially convenient case is when the beta process is homogeneous, i.e., $c(\omega) = c$ is a constant. In this case $\{\omega_{ki}\}_{i=1}^{n_k}$ for all rounds $k$ are drawn from the same distribution $\mu/\gamma$; and $n_k$ is drawn from $\text{Poisson}(\frac{c\gamma}{c+k})$. For round $k$, both the number of points and the jumps statistically diminish as $k$ increases, suggesting that the infinite sum in (9) may be truncated as $B = \sum_{k=0}^{K} B_k$ for large $K$, with minimal impact. Such truncation effects are investigated in detail in Section 3.4.

#### 3.3.2. Related work

In (Thibaux & Jordan, 2007) the authors derived the above simulation procedure for the homogeneous case within the beta-Bernoulli process context, which is shown here a necessary result of the Lévy measure decomposition. The same decomposing manipulation of Theorem 1 can be also applied to the stable beta process (Teh & Görür, 2009) which yields:

$$\nu_k = \text{Beta}(1 - \sigma, c(\omega) + \sigma + k) d\pi \\ \cdot \frac{\Gamma(c(\omega) + \sigma + k)\Gamma(c(\omega) + 1)}{\Gamma(c(\omega) + k + 1)\Gamma(c(\omega) + \sigma)} \mu(d\omega) \quad (13)$$

It is noteworthy that the decomposition procedure described in Theorem 1 is not the only Lévy measure decomposing method for the beta process. The work of (Paisley & Jordan, 2012) and (Broderick et al., 2011) show that the stick-breaking construction of the beta process in (Paisley et al., 2010) is indeed a result of another way of decomposing the Lévy measure of the beta process. We next analyze the truncation property of the construction described in Section 3.3.1 and make comparison with the construction of beta process in (Paisley et al., 2010).

### 3.4. Truncation analysis

Since the Poisson superposition simulation operates in rounds, it is natural to analyze the distance between the true beta process $B$ and its truncation $\sum_{k=0}^{K} B_k$, with truncation at round $K$. A metric for such distance is the $\mathcal{L}_1$ norm:

$$||B - \sum_{k=0}^{K} B_k||_1 = \mathbb{E}|B - \sum_{k=0}^{K} B_k| = \int_\Omega \frac{\mu_{K+1}(d\omega)}{\gamma} \quad (14)$$

The expectation in (14) is w.r.t. the normalized measure $\nu/\gamma$, which yields $\|B\|_1 = 1$. When $B$ is homogeneous, (14) reduces to $\frac{c}{c+K+1}$, which indicates that the $\mathcal{L}_1$ distance decreases at a rate of $\mathcal{O}(\frac{1}{K})$. For the stick-breaking construction of beta process described in (Paisley et al., 2010), the $\mathcal{L}_1$ distance is: $(\frac{c}{c+1})^{K+1}$.

Another metric is the $\mathcal{L}_1$ distance between the marginal likelihood of a set of data $\boldsymbol{b} = b_{1:M}$, with $m_\infty(\boldsymbol{b})$ denotes the marginal likelihood (here the likelihood is a Bernoulli process) with prior $B$, and $m_K(\boldsymbol{b})$ for $\sum_{k=0}^{K} B_k$. This metric was applied on the truncated Indian buffet process (Doshi et al., 2009) and truncated stick-breaking construction of the beta process (Paisley & Jordan, 2012), which indicates

$$\frac{1}{4} \int |m_\infty(\boldsymbol{b}) - m_K(\boldsymbol{b})| d\boldsymbol{b} \leq \\ \Pr(\exists k > K, \ 1 \leq i \leq n_k, \ 1 \leq m \leq M, \ \text{s.t.} \ b_{ki}^m = 1) \quad (15)$$

where $b_{1:M} \overset{\text{i.i.d.}}{\sim} \text{BeP}(B)$ are drawn from a Bernoulli process with base measure $B$; $b_{ki}^m = b_m(\omega_{ki})$ is the $m^{th}$ realization of the Bernoulli process at atom $\omega_{ki}$. For the truncation $\sum_{k=0}^{K} B_k$ it can be shown that the RHS of (15) is bounded by:

$$\text{RHS of (15)} \leq 1 - \exp(-M \int_\Omega \mu_{K+1}(d\omega)) \quad (16)$$

For the homogeneous case, the bound of (16) is $1 - \exp(-M\gamma \frac{c}{c+K+1})$. For the stick-breaking construction of beta process, the bound is given by: $1 - \exp(-M\gamma(\frac{c}{c+1})^{K+1})$ (Paisley & Jordan, 2012).

In order to analyze the bound w.r.t. the truncation level by number of atoms, denote $I_K = \sum_{k=0}^{K} n_k$ as the total number of atoms in $\sum_{k=0}^{K} B_k$. Since $K \sim \mathcal{O}(e^{\frac{\mathbb{E}(I_K)}{c\gamma}})$, it is proved that (14) and the bound in (16) decreases at a faster rate w.r.t. $I$ than the stick-breaking construction of beta process. This indicates that the simulation procedure described in Section 3.3.1 follows a steeper statistically-decreasing order. The proof is presented in the Supplementary Material.

### 3.5. Posterior estimation

The goal of the inference is to estimate the beta process $B$ from a set of observed data $\boldsymbol{b}$ with prior $\text{BP}(c, \mu)$. The data $\boldsymbol{b} = b_{1:M}$ is the same as in Section 3.4, which can be expressed as:

$$b_m = \sum_{i=1}^{\infty} b_{i,m} \delta_{\omega_i}, \quad m = 1, 2, \cdots, M \quad (17)$$

where each $b_{i,m} \in \{0, 1\}$.

#### 3.5.1. POSTERIOR OF $B_k$

Since $B|\boldsymbol{b} \sim \text{BP}(c+M, \frac{c\mu}{c+M} + \frac{\sum_{m=1}^{M} b_m}{c+M})$ (Thibaux & Jordan, 2007), the base measure of $B|\boldsymbol{b}$ is a measure with positive masses assigned on single atoms. Theorem 1 is still applicable to this beta process with mixed type of base measure, which yields

$$B' = \sum_{k=0}^{\infty} B'_k$$
$$\nu'_k = \text{Beta}(1, c+M+k)\mu'_k \quad (18)$$
$$\mu'_k = \frac{c\mu}{c+M+k} + \frac{\sum_{m=1}^{M} b_m}{c+M+k}$$

where the $B'$, $B'_k$, $\nu'_k$, and $\mu'_k$ are the posterior counterparts of $B$, $B_k$, $\nu_k$, and $\mu_k$.

#### 3.5.2. POSTERIOR ESTIMATION OF $\pi_i$:

Since each $\mu_k$ has a mass $\frac{\sum_{m=1}^{M} b_{i,m}}{c+M+k}$ at the atom $\omega_i$, each $B_k$ will contribute $\text{Poisson}(\frac{\sum_{m=1}^{M} b_{i,m}}{c+M+k})$ draws with the jumps following the distribution $\text{Beta}(1, c+M+k)$ at the atom $\omega_i$, whose sum is the $\pi_i$. Thus the posterior estimation of $\pi_i$ is given by

$$\pi_i|\boldsymbol{b} = \sum_{k=0}^{\infty} \sum_{h=1}^{H_k} b_{kh}$$
$$H_k \sim \text{Poisson}(\frac{\sum_{m=1}^{M} b_{i,m}}{c+M+k}) \quad (19)$$
$$b_{kh} \sim \text{Beta}(1, c+M+k)$$

from which it can be verified that $\mathbb{E}(\pi_i|\boldsymbol{b}) = \frac{\sum_{m=1}^{M} b_{i,m}}{c+M}$, the same as the posterior of $\pi_i$ without decomposition: $\text{Beta}(\sum_{m=1}^{M} b_{i,m}, c+M-\sum_{m=1}^{M} b_{i,m})$.

For the $\pi_i$ with no observations, i.e., $\sum_{m=1}^{M} b_{i,m} = 0$, only a particular $B_k$ will contribute to $\pi_i$. In this case, first the round $k$ to which $\pi_i$ belongs is drawn, then $\pi_i$ is drawn from the beta distribution of that round:

$$\pi_i \sim \text{Beta}(1, c+M+k)$$
$$k \sim \text{MP}(\boldsymbol{\alpha}), \quad \boldsymbol{\alpha} \propto \sum_{k=0}^{\infty} \frac{1}{c+M+k} \delta_k \quad (20)$$

where $\text{MP}(\boldsymbol{\alpha})$ is a multinomial process with probability vector $\boldsymbol{\alpha}$, and $\boldsymbol{\alpha}$ is proportional to the average number of points in each round. Since in practical processing $\boldsymbol{\alpha}$ is always to be truncated with a truncation level $K$, by the analysis in Section 3.4, (20) provides a way to estimate the $\pi_i$ within the first $K$ rounds. And $\pi_i$ in each round are of statistically different importance, contrasted to the evenly assigned mass in the Indian buffet process.

### 3.6. Relating the IBP and beta process

The study of the beta process through its Lévy measure, as discussed in this paper, also uncovers a connection between the Indian buffet process (IBP) (Griffiths & Ghahramani, 2005) and the beta process, by their Lévy measures. The IBP with prior $\pi_i \sim \text{Beta}(c\frac{\gamma}{N}, c)$ can be regarded as a Lévy process with the Lévy measure given as:

$$\nu_{\text{IBP}} = \frac{N}{\gamma} \text{Beta}(c\frac{\gamma}{N}, c) d\pi \mu(d\omega) \quad (21)$$

here $N$ is the same as the $K$ in (Griffiths & Ghahramani, 2005). It can be proved that:

$$\nu_{\text{IBP}} \stackrel{N \to \infty}{=} \nu \quad (22)$$

which indicates that the beta process is the limit of the IBP with $N \to \infty$. The detailed proof of (22) is presented in the Supplementary Material. Thus the IBP is like a "mosaic" approximation of beta process, which becomes finer with $N$ increases.

## 4. Gamma process

A gamma process (Applebaum, 2009) is a Lévy process with independent gamma increments. The gamma process is traditionally parameterized with a shape measure and a scale function: $G \sim \Gamma\mathrm{P}(\alpha, \theta(\omega))$ where $\alpha$ is the shape measure on a measure space $(\Omega, \mathcal{F})$, and the scale $\theta(\omega)$ a positive function. A gamma process can be intuitively defined by its increments on infinitesimal sets:

$$G(d\omega) \sim \mathrm{Gamma}(\alpha(d\omega), \theta(\omega)) \tag{23}$$

When $\theta(\omega) = \theta$ is a scalar, the gamma process is called homogeneous. The gamma process can also be expressed in the form with a base measure $G_0$ and a concentration $c(\omega)$, with $c = 1/\theta$ and $G_0 = \theta\alpha$ (Jordan., 2009), to conform with other stochastic processes widely used in machine learning, such as the Dirichlet process. However, the discussion in this paper will stick to the traditional form given by (23).

As a pure-jump Lévy process, the gamma process can be regarded as a Poisson process on the product space $\Omega \times \mathbb{R}^+$ with mean measure $\nu$:

$$\nu(dp, d\omega) = p^{-1}\mathrm{e}^{-\frac{p}{\theta(\omega)}} dp\alpha(d\omega) \tag{24}$$

where $\mathrm{Gamma}(0, \theta(\omega)) = p^{-1}\mathrm{e}^{-\frac{p}{\theta(\omega)}}$ is an improper gamma distribution with an infinite integral on $\mathbb{R}^+$, which yields the expression of $G$:

$$G = \sum_{i=1}^{\infty} p_i \delta_{\omega_i} \tag{25}$$

### 4.1. Lévy measure decomposition

Like the beta process, the Lévy measure of the gamma process is characterized by an improper distribution. However, unlike the beta process, the decomposition of the Lévy measure of the gamma process comes from the exponential part. With the details shown in the Supplementary Material, the gamma process $G$ can be decomposed into two parts:

$$G = \Gamma_1 + \Gamma\mathrm{P}(\alpha, \theta(\omega)/2) \tag{26}$$

The second term in (26) is a gamma process with the same shape measure, and half the scale of the gamma process $G$; the first term $\Gamma_1$ is a Lévy process with the Lévy measure $\sum_{h=1}^{\infty} \mathrm{Gamma}(h, \frac{\theta(\omega)}{2})dp\frac{\alpha(d\omega)}{2^h h}$. Here $\mathrm{Gamma}(h, \frac{\theta(\omega)}{2})$ is the PDF of the gamma distribution, with shape parameter $h$ and scale parameter $\frac{\theta(\omega)}{2}$.

Further decomposing the exponential part of the gamma process $\Gamma\mathrm{P}(\alpha, \theta(\omega)/2)$ in (26) yields $G = \Gamma_1 + \Gamma_2 + \Gamma\mathrm{P}(\alpha, \theta(\omega)/3)$, bearing a gamma process with the same shape and with the scale parameter further decreased. Repeating this manipulation, we obtain the Theorem 2:

**Theorem 2** *A gamma process $G \sim \Gamma\mathrm{P}(\alpha, \theta(\omega))$ with shape measure $\alpha$ and scale $\theta(\omega)$ can be decomposed as:*

$$\begin{aligned} G = \sum_{k=1}^{\infty} \Gamma_k, \ \Gamma_k = \sum_{h=1}^{\infty} \Gamma_{kh}, \ \nu_k = \sum_{h=1}^{\infty} \nu_{kh} \\ \nu_{kh} = \mathrm{Gamma}(h, \frac{\theta(\omega)}{k+1})dp\frac{\alpha(d\omega)}{(k+1)^h h} \end{aligned} \tag{27}$$

*with $\Gamma_k$, $\Gamma_{kh}$ Lévy processes with $\nu_k$, $\nu_{kh}$ their Lévy measures.*

Theorem 2 is the gamma process instantiation of (4), which indicates that $G$ can be expressed as the sum of an infinite number of Lévy processes $\Gamma_k, k = 1, 2, \cdots$, where $\Gamma_k$ is also the sum of an infinite number of Lévy processes $\Gamma_{kh}, h = 1, 2, \cdots$.

### 4.2. Lévy processes $\Gamma_k$ and $\Gamma_{kh}$

In order to obtain further insights into the gamma process $G$ in Theorem 2, the expectations and variances of $\Gamma_k$ and $\Gamma_{kh}$ on any measurable set $\mathcal{A} \in \mathcal{F}$ are given:

$$\begin{aligned} \mathbb{E}(\Gamma_{kh}(\mathcal{A})) = \frac{\int_{\mathcal{A}} \theta(\omega)\alpha(d\omega)}{(k+1)^{h+1}} \\ \mathbb{E}(\Gamma_k(\mathcal{A})) = \frac{\int_{\mathcal{A}} \theta(\omega)\alpha(d\omega)}{k(k+1)} \end{aligned} \tag{28}$$

For the variances of $\Gamma_k$ and $\Gamma_{kh}$:

$$\begin{aligned} \mathrm{Var}(\Gamma_{kh}(\mathcal{A})) = \frac{(h+1)}{(k+1)^{h+2}} \int_{\mathcal{A}} \theta^2(\omega)\alpha(d\omega) \\ \mathrm{Var}(\Gamma_k(\mathcal{A})) = [\frac{1}{k^2} - \frac{1}{(k+1)^2}] \int_{\mathcal{A}} \theta^2(\omega)\alpha(d\omega) \end{aligned} \tag{29}$$

Since the Lévy processes $\Gamma_k$ are independent w.r.t. $k$, with analogy to (12) it can be verified that the expectation and variance of $\Gamma_k$ sum to the expectation of variance of $G$. The derivations in this section are presented in the Supplementary Material.

### 4.3. Simulation of gamma process

Parallel to the simulation of beta process in Section 3.3.1, a simulation procedure of the gamma process is presented:

**Simulation procedure**: Sample the Lévy process $\Gamma_{kh}$:

1: Sample the number of points for $\Gamma_{kh}$: $n_{kh} \sim$ Poisson$(\gamma/(k+1)^h h)$;
2: Sample $n_{kh}$ points from $\alpha$: $\omega_{khi} \overset{\text{i.i.d.}}{\sim} \frac{\alpha}{\gamma}$, for $i = 1, 2, \cdots, n_{kh}$;
3: Sample $\Gamma_{kh}(\omega_{khi}) \overset{\text{i.i.d.}}{\sim}$ Gamma$(h, \frac{\theta(\omega_{khi})}{k+1})$, for $i = 1, 2, \cdots, n_{kh}$;

where $\gamma = \int_\Omega \alpha(d\omega)$ is the mass of the shape measure. Then the union $\bigcup_{k=1}^\infty \bigcup_{h=1}^\infty (\omega_{khi}, \Gamma_{kh}(\omega_{khi}))_{i=1}^{n_{kh}}$ is a realization of the gamma process $G$. An advantage of the above simulation procedure compared to the simulation procedure of the beta process in Section 3.3.1 is that independent of whether the gamma process is homogeneous or inhomogeneous, $\omega_{khi}$ is always drawn from a fixed distribution $\alpha/\gamma$. Like with the beta process construction in Section 3.3.1, for the gamma process simulation procedure, as $k$ increases the expected number of new points and the expected jumps decrease, again suggesting accurate truncation.

### 4.4. Truncation analysis

Since in the simulation procedure in Section 4.3 the index $k$ and $h$ both go to infinity, it is practical to analyze the distance between the true gamma process and the truncated one. To measure such a distance, we apply the $\mathcal{L}_1$ norm described in Section 3.4:

$$||G - \sum_{k=1}^K \sum_{h=1}^H \Gamma_{kh}||_1 = \mathbb{E}|G - \sum_{k=1}^K \sum_{h=1}^H \Gamma_{kh}| \quad (30)$$

where the expectation in (30) is w.r.t. the normalized measure $\nu / \int_\Omega \theta(\omega) \alpha(d\omega)$ with $||G||_1 = 1$; and $K$ and $H$ are the truncation level of $k$ and $h$. Then for the situation with $H = \infty$:

$$||G - \sum_{k=1}^K \sum_{h=1}^\infty \Gamma_{kh}||_1 = \frac{1}{K+1} \quad (31)$$

which indicates a $\mathcal{O}(\frac{1}{K})$ decreasing rate as same as the truncated beta process shown in (14). It is noteworthy that $\Gamma_1$ alone accounts for on average half the mass of $G$. When $H$ is finite, a remaining distance $\sum_{k=1}^K \frac{1}{k(k+1)^{H+1}}$ is added.

### 4.5. Generalized gamma process and symmetric gamma process

Theorem 2 can be easily extended to some variations of the gamma process. Here we give the examples of the generalized gamma process (Brix, 1999) and symmetric gamma process (Çinlar, 2010).

The generalized gamma process extends the ordinary gamma process by adding a parameter $0 < \sigma < 1$, whose Lévy measure is $\frac{1}{\Gamma(1-\sigma)} p^{-\sigma-1} e^{-\frac{p}{\theta(\omega)}} dp \alpha(d\omega)$. Then with the same decomposition procedure, it is straightforward that the Lévy measure for $\Gamma_{kh}$ of the generalized gamma process will change to $\nu_{kh} =$ Gamma$(h - \sigma, \frac{\theta(\omega)}{k+1}) dp \frac{\alpha(d\omega)}{\Gamma(1-\sigma)(k+1)^h h}$.

The symmetric gamma process is a Lévy process whose increments are the differences of two gamma-distributed variables with the same law, whose Lévy measure is $|p|^{-1} e^{-\frac{|p|}{\theta(\omega)}} dp \alpha(d\omega)$. Since there can be negative increments, the symmetric gamma process is not a completely random measure. However, the same decomposition procedure is still applicable, yielding $\nu_{kh} =$ Gamma$(|p||h, \frac{\theta(\omega)}{k+1}) dp \frac{2\alpha(d\omega)}{(k+1)^h h}$, where the distribution Gamma$(|p||h, \frac{\theta(\omega)}{k+1})$ is to first draw $|p|$ from Gamma$(h, \frac{\theta(\omega)}{k+1})$, then decide the sign of $p$ through a symmetric Bernoulli distribution.

## 5. Conclusions

The Lévy measure decomposition of the beta and gamma processes provides new perspectives on the two widely used stochastic processes, by casting insights on the sub-processes constituting them, here the $B_k$ and $\Gamma_k$. And the decomposition prescriptions described here are far from the only ways of such decomposition. Theoretically elegant construction methods are derived from the proposed decompositions, which are directly implementable in practice.

We have applied the proposed beta and gamma representations in numerical experiments, the details of which are omitted, as this paper focuses on foundational properties. However, to briefly summarize experience with such representations, consider for example the image inpainting problem considered in (Zhou et al., 2009), based upon a beta process factor analysis model (Paisley & Carin, 2009). In experiments we performed with such a model, using a Gibbs sampler, the beta process prior was implemented using the procedure discussed in Section 3.3.1, with the posterior estimation in Section 3.5 applied for inference. The proposed representation infers a dictionary with the "important" dictionary elements captured by the low-index members (see the discussion in Section 3.3.1). The model prioritized the first three dictionary elements as being pure colors, specifically red, green, and blue, with the important structured dictionary elements following (and no other pure-color dictionary elements, while in (Zhou et al., 2009) many – seemingly redundant – pure-color dictionary elements are inferred). This "clean" inference of prioritized dictionary elements may be responsible for our also higher observed PSNR in signal recovery, compared to the re-sult given in (Zhou et al., 2009). The new gamma process construction in Section 4.3 may be implemented in a similar manner, and may be employed within recent models in machine learning in which the gamma process has been utilized (*e.g.*, (Paisley et al., 2011)).

## Acknowledgements

The research reported here was supported by ARO, NGA, ONR and DARPA (MSEE program).